\documentclass{PoS}

\usepackage{amsmath}
\usepackage{microtype}

\title{Probing nuclear PDFs with dijets in
ultra-peripheral Pb+Pb collisions}

\ShortTitle{Probing nuclear PDFs with dijets in
ultra-peripheral Pb+Pb collisions}

\author{\speaker{Ilkka Helenius}\\
        University of Jyvaskyla, Department of Physics, P.O. Box 35, FI-40014 University of Jyvaskyla, Finland \\
        Institute for Theoretical Physics, T\"ubingen University, Auf der Morgenstelle 14, 72076 T\"ubingen, Germany \\
        E-mail: \email{ilkka.m.helenius@jyu.fi}}

\abstract{%Precision of the current nuclear PDF (nPDF) analyses is limited due to the lack of data constraints especially at small-x region. Ultimately the best way to pin down the nPDF uncertainties would be the clean photon-induced processes in a high-energy lepton-ion collider. To some extend this can also be accomplished with ultra-peripheral heavy-ion collisions at the LHC where a quasi-real photon from one nucleus interacts with a nucleus from the other beam in an event where a large impact parameter prevents hadronic interactions.
In this talk we apply the photoproduction framework recently implemented into the \textsc{Pythia}~8 Monte Carlo event generator to study the potential of photo-nuclear dijets in ultra-peripheral Pb+Pb collisions at the LHC to further constrain the nuclear PDFs. These events can be described as $\gamma$A collisions where the relevant part of the flux of quasi-real photons from heavy-ions is obtained by using the equivalent photon approximation and cutting out impact-parameter values which would lead to hadronic interactions between the beam particles. In particular, we quantify the small-$x$ reach with different jet kinematics and show how well the values of $x$ derived from reconstructed jet momenta are correlated with the actual values of partonic momentum fractions probed in these measurements. Also the contributions from direct and resolved photons are separately presented. To demonstrate the potential, we compare the expected experimental uncertainties to the current nuclear-PDF errors and discuss other theoretical uncertainties including the uncertainty arising from poorly-constrained photon PDFs. We find that such a measurement would potentially provide a considerable reduction of the nuclear PDF uncertainties in a region $10^{-4} \lesssim x \lesssim 10^{-2}$.
}

\FullConference{International Conference on Hard and Electromagnetic Probes of High-Energy Nuclear Collisions\\
		30 September - 5 October 2018\\
		Aix-Les-Bains, Savoie, France}

\begin{document}

\section{Introduction}

The uncertainties in the current nuclear PDF (nPDF) analyses \cite{Eskola:2016oht, Kovarik:2015cma} are considerable especially at small values of $x$ due to lack of data constraints. 
%The current nuclear PDF (nPDF) analyses \cite{Eskola:2016oht, Kovarik:2015cma} are not well-constrained by the considerable especially at small values of $x$ due to lack of data constraints. 
%Due to limited amount of available data the current nuclear PDF (nPDF) analyses \cite{Eskola:2016oht, Kovarik:2015cma} have large uncertainties especially at small values of $x$.
The most precise way to pin down the nPDF uncertainties would be to measure the structure functions in deep inelastic scattering (DIS) process in a high-energy electron-ion collider where a virtual photon probes the partonic structure of a heavy nucleus \cite{Aschenauer:2017oxs}. Such colliders are being considered (see e.g. Ref.~\cite{Accardi:2012qut}) but to some extent similar measurements could be performed also in ultra-peripheral Pb+Pb collisions at the LHC. 

In ultra-peripheral collisions (UPCs) \cite{Baltz:2007kq} the beam particles pass without any strong interactions between the constituting partons but a quasi-real photon emitted by one beam particle collide with a particle from the other beam or with a photon from the other beam. Unlike in DIS processes the virtuality of these intermediate photons is small so the hard scale justifying the use of perturbative QCD must be provided by a high-$p_{\mathrm{T}}$ observable. Furthermore, these quasi-real photons may fluctuate into a hadronic state with the same quantum numbers. The partonic structure of these resolved photons can be described with PDFs and, similarly as in the case of purely hadronic collisions, gives rise to multiple partonic interactions (MPIs). Due to the inevitable contribution of resolved photons, the measurements in this low-$Q^2$ (photoproduction) regime are not as clean as in high-$Q^2$ (DIS) but due to presence of the direct-photon -parton scattering, and different partonic structure, the underlying event is significantly suppressed compared to hadronic collisions.

In this work we study the possibility to further constrain the nPDFs using dijet production in UPCs at the LHC as proposed in Ref.~\cite{Strikman:2005yv}. The first experimental results for the process have been recently published by the ATLAS collaboration \cite{ATLAS:2017kwa}. Dijet production in p+Pb collisions at the LHC has already been considered as a probe for nPDFs (see e.g. Ref.~\cite{Eskola:2013aya}) but the UPC measurement would allow to study jet production in a cleaner $\gamma A$ interaction that should enable jet reconstruction at lower values of $p_{\mathrm{T}}$. This increases the small-$x$ reach of the observable and also lowers the scale at which the nPDFs are probed where the uncertainties are larger. In this study the dijet cross sections are calculated applying the recent photoproduction framework of \textsc{Pythia 8} \cite{Sjostrand:2014zea} Monte Carlo event generator that has been validated against various HERA data \cite{Helenius:2017aqz}.

\section{Ultra-peripheral Pb+Pb collisions at the LHC}

\begin{figure}[htb!]
\centering
\includegraphics[width=0.45\textwidth]{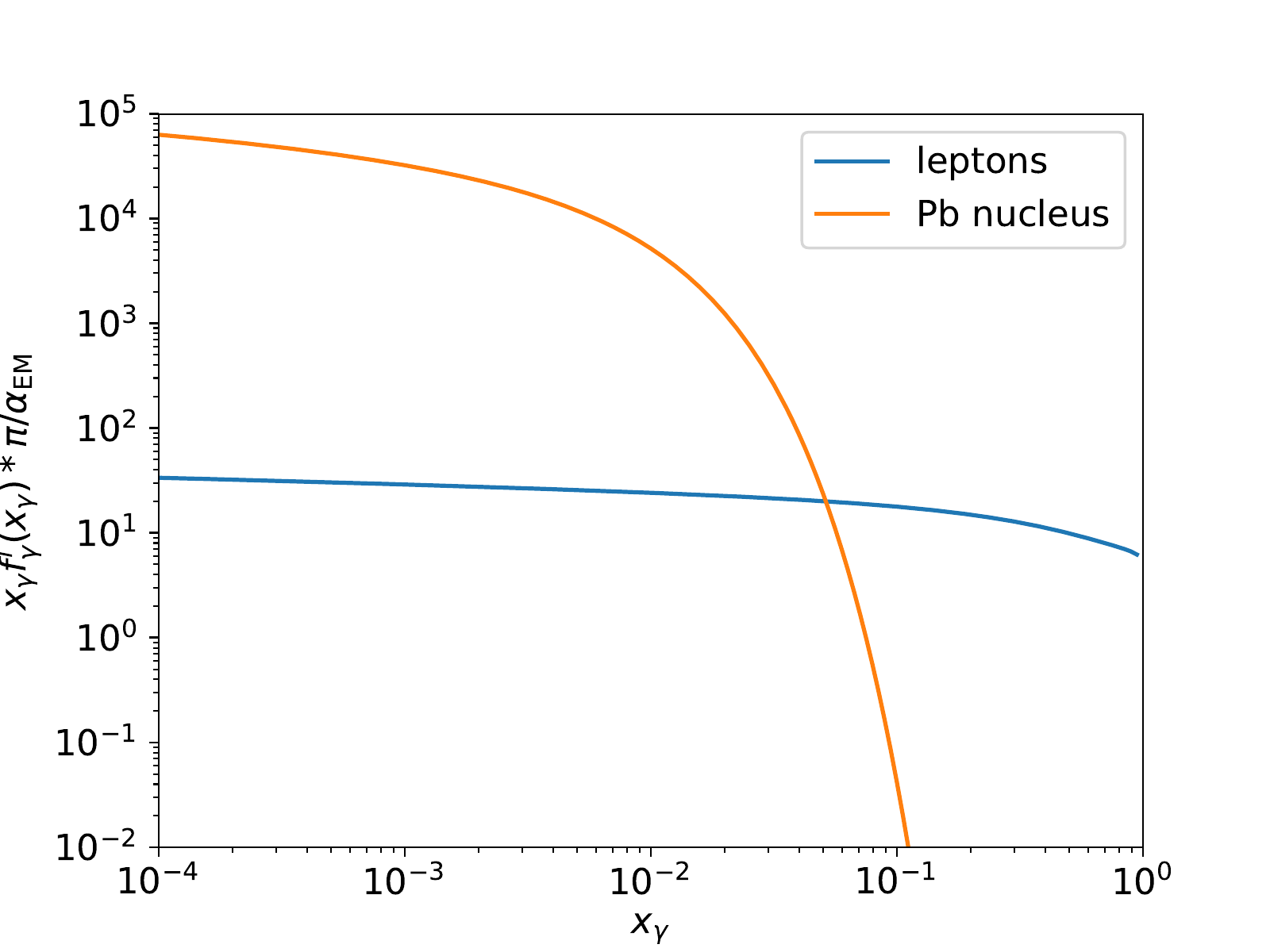}
\includegraphics[width=0.45\textwidth]{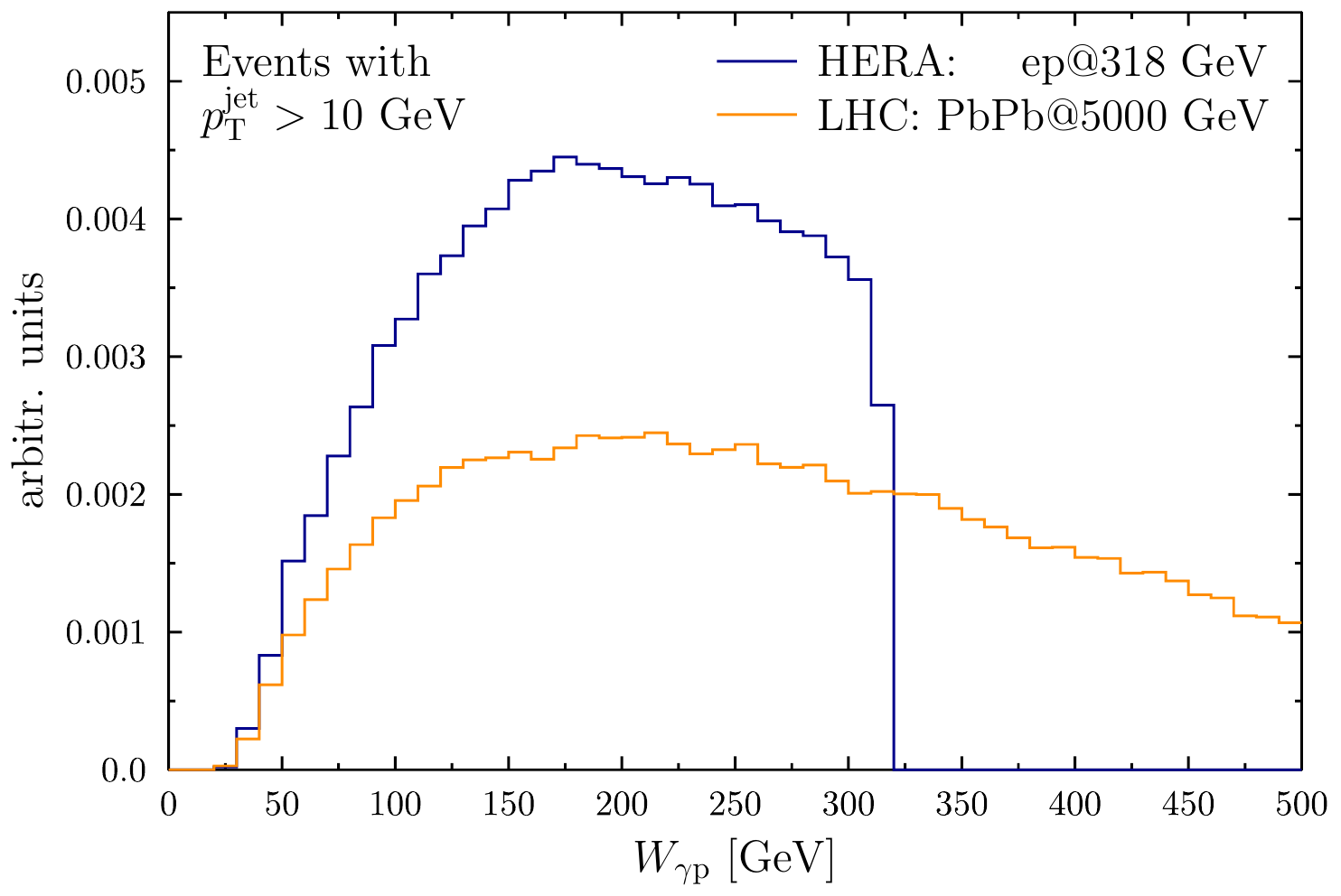}
\caption{\textbf{Left:} The photon flux from leptons (orange) and from Pb-nucleus (blue) as a function of photon momentum fraction $x_{\gamma}$. \textbf{Right:} Invariant-mass distributions of $\gamma$-nucleon system in e+p collisions at HERA (orange) and ultra-peripheral Pb+Pb collisions at the LHC (blue) for events that have a partonic jet with $p_{\mathrm{T}}>10~\mathrm{GeV}/c$.}
\label{fig:fluxComp}
\end{figure}

The photon flux from heavy ions can be obtained according to the equivalent photon approximation. Since the configurations where the two ions would interact hadronically must be rejected, it is convenient to define the photon flux in impact-parameter ($b$) space. To a good precision, the relevant part of the photon flux can be calculated by integrating the flux over impact parameters larger than the sum of radii of the colliding ions ($R_A$), giving
\begin{equation}
x_{\gamma}f_{\gamma}^{A}(x_{\gamma}) = \frac{2\alpha_{\mathrm{EM}}Z^2}{\pi}\left[ \xi K_1(\xi)K_0(\xi) - \frac{\xi^2}{2}\left( K_1^2(\xi) - K_0^2(\xi) \right) \right],
\end{equation}
where $x_{\gamma}$ is the momentum fraction of the photon wrt. the beam particle, $\xi = 2 R_A x_{\gamma} m / (\hbar c)$, $m$ (per-nucleon) mass and $Z$ the electric charge of the nucleus. Compared to the flux of photons from leptons the flux from heavy ions is amplified by a factor of $Z^2$ but, due to the finite size of the nuclei taken into account with the impact-parameter based rejection, the photon spectrum is softer as shown in Fig.~\ref{fig:fluxComp}. The invariant mass distributions of the $\gamma$-nucleon system, $W_{\gamma \mathrm{p}}$, for events that have a partonic jet with $p_{\mathrm{T}}>10~\mathrm{GeV}/c$ is shown in Fig.~\ref{fig:fluxComp} for setups corresponding to the photoproduction measurements at HERA and the UPCs at the LHC. As the photon spectrum from nuclei is rather soft, large contribution of the dijet cross section in UPCs is coming from $W_{\gamma \mathrm{p}}$ values probed at HERA but due to higher $\sqrt{s_{\mathrm{NN}}}$ at the LHC, the $W_{\gamma \mathrm{p}}$ distribution extends still up to a factor of two higher energies than what was reached at HERA. For the UPC events the parton showers and MPIs are generated for the $\gamma$-nucleon system so further scatterings of the photon remnant with the other nucleons are not considered.

\section{Dijet sensitivity to nPDFs}

\begin{figure}[thb!]
\centering
\includegraphics[height=165pt]{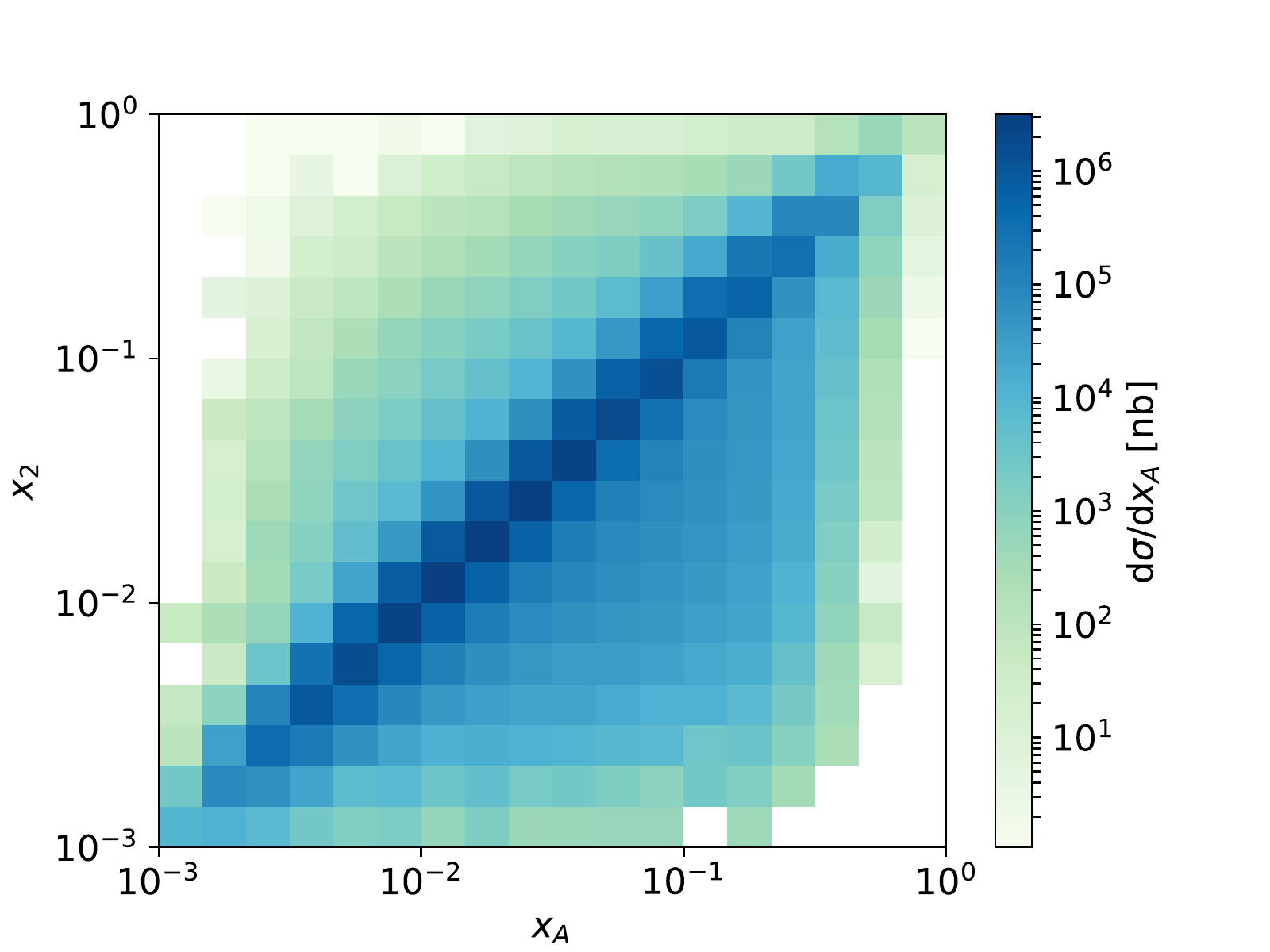}
\includegraphics[height=165pt]{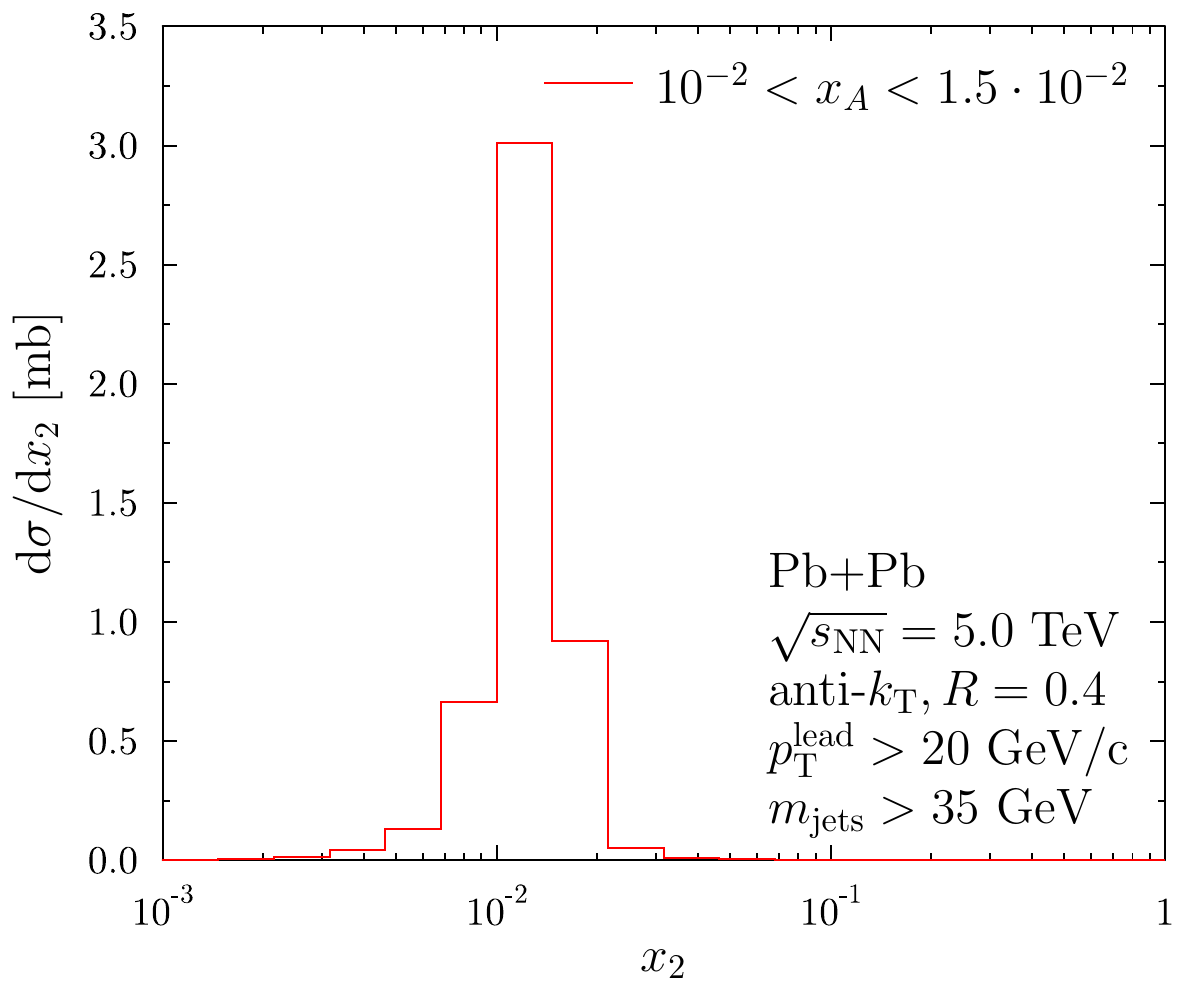}
\caption{\textbf{Left:} Correlation between $x_A$ and the actual $x_2$ at which the nPDFs are probed with dijets in the ATLAS kinematics. \textbf{Right:} Cross section for events within a given $x_A$ bin as a function of $x_2$.}
\label{fig:correlation}
\end{figure}

We study the photo-nuclear dijet production in two different jet kinematics. First is based on the preliminary ATLAS study \cite{ATLAS:2017kwa} where the $p_{\mathrm{T}}$ of the leading jet was above $20~\mathrm{GeV}/c$ and the sub-leading jets had $p_{\mathrm{T}}>15~\mathrm{GeV}/c$. In the second case the corresponding limits for the $p_{\mathrm{T}}$ of the leading and sub-leading jets are $8~\mathrm{GeV}/c$ and $6~\mathrm{GeV}/c$ which resembles the jet kinematics analyzed at HERA \cite{Adloff:2000bs}. In both cases the jets are reconstructed with the anti-$k_{\mathrm{T}}$ algorithm with $R=0.4$ and are required to be within $|\eta|<4.4$ and events with at least two accepted jets are considered. As in the ATLAS study we construct the following event-level variables from the four momenta of the reconstructed jets
\vspace{-0.3cm}
\begin{align*}
m_{\mathrm{jets}} &= \sqrt{ \left(\Sigma_i E_i\right)^2 - \left| \Sigma_i \vec{p}_i \right|^2 } \hspace{2.9em} H_{\mathrm{T}}= \Sigma_{i} p_{\mathrm{T}i}  \\
y_{\mathrm{jets}} &= \frac{1}{2} \log \left( \frac{\Sigma_i E_i + \Sigma_i p_{zi}}{\Sigma_i E_i - \Sigma_i p_{zi}} \right) \hspace{2em} x_A = \frac{m_{\mathrm{jets}}}{\sqrt{s}}\mathsf{e}^{-y_{\mathrm{jets}}},
\end{align*}
where the sums run over accepted jets. For the nPDF studies particularly interesting observable is the $x_A$ variable, which, in leading order parton-level kinematics would correspond to $x_2$, the momentum fraction of the parton in the nucleus. However, this connection is smeared due to parton-shower emissions, hadronization effects and underlying event from the MPIs. To study the correlation between $x_A$ and $x_2$ within the applied MC framework we plot the two-dimensional cross section in terms of these variables in Fig.~\ref{fig:correlation}. For a more quantitative result we also plot the cross section as a function of $x_2$ in one $x_A$ bin in linear scale in Fig.~\ref{fig:correlation}. The results show a strong correlation between $x_A$ and $x_2$.
% which enables to measure the nPDFs 
% to use this observable to constrain the nPDFs at 

\begin{figure}[htb!]
\centering
\includegraphics[width=0.45\textwidth]{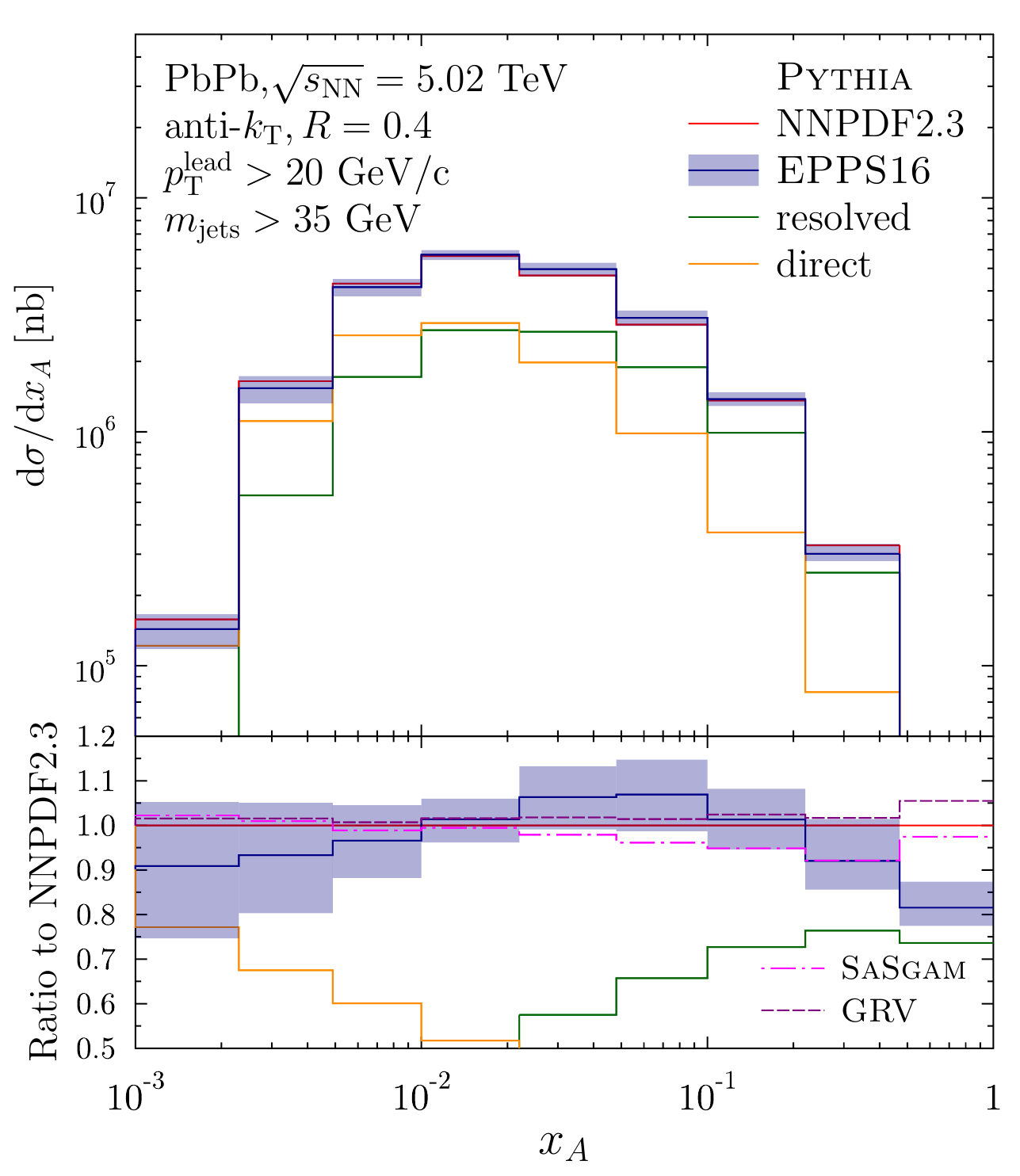}
\includegraphics[width=0.45\textwidth]{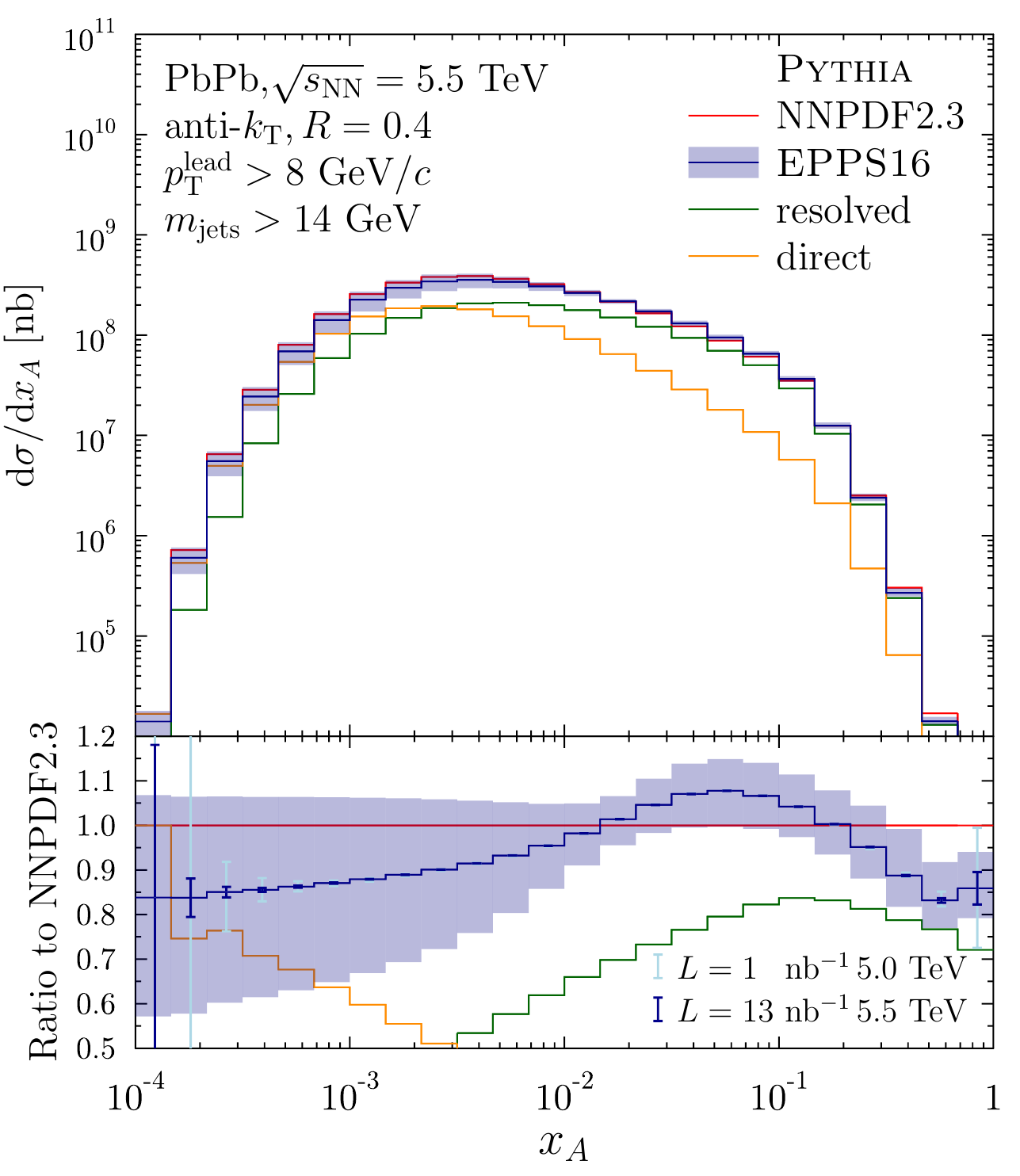}
\caption{Differential cross section of photo-nuclear dijet production in ultra-peripheral Pb+Pb collisions at the LHC as a function of $x_A$. Results with (blue) and without (red) EPPS16 nPDFs are shown separately as also the contributions from direct (orange) and resolved processes (green). The lower panels show ratios to the result without the EPPS16 nPDFs. With the higher $p_{\mathrm{T}}$ cuts (left) also the ratios to different photon PDFs are shown and the expected statistical uncertainties corresponding to the integrated luminosities of the LHC and HL-LHC are estimated for lower $p_{\mathrm{T}}$ cuts (right).}
\label{fig:CrossSect}
\end{figure}

To calculate the dijet cross section we apply NNPDF2.3 LO \cite{Ball:2012cx} proton PDFs with EPPS16 NLO \cite{Eskola:2016oht} nuclear modifications. The resulting cross sections corresponding to the two different cuts for jet $p_{\mathrm{T}}$'s are presented in Fig.~\ref{fig:CrossSect} as a function of $x_A$. The resolved-photon contribution dominates the cross section at large-$x_A$ region but the direct contribution takes over at around $x_A<0.01\,(0.002)$ with the harder (softer) jet cuts. The impact of nPDFs is quantified by showing a ratio to the calculation with the proton PDFs only. The nPDF-originating uncertainties from EPPS16 analysis are of the order 10\% at $x_A>0.01$ but grow up to 30\% at smaller values of $x_A$ with the softer jet cuts. By default \textsc{Pythia}~8 takes the PDFs for the resolved photons from CJKL analysis \cite{Cornet:2002iy} but we have considered also two other widely-used sets, GRV \cite{Gluck:1991jc} and SaSgam (1D V2) \cite{Schuler:1995fk} to study the uncertainties related to precision of the photon PDFs. The resulting uncertainty is up to 10\% at the highest values of $x_A$ where the resolved contribution dominates but is negligible at smaller values of $x_A$ where the nuclear PDF uncertainties are large. In Fig.~\ref{fig:CrossSect} we also estimate the expected statistical precision for the softer $p_{\mathrm{T}}$ jets assumming integrated luminosity of $L=1.0~\mathrm{nb}^{-1}$ corresponding to the heavy-ion program at the LHC and $L=13~\mathrm{nb}^{-1}$ corresponding to the high-luminosity LHC (HL-LHC) proposal. The expected statistical uncertainties become larger than the nPDF uncertainties at $x_A \lesssim 2\cdot 10^{-4}$.

%Compared to the current nPDF uncertainties these become dominant only at $x_A \lesssim 2\cdot 10^{-4}$ demonstrating the potential of such a measurement.

\section{Conclusions}

We have studied the photo-nuclear dijet production in ultra-peripheral Pb+Pb collisions at the LHC and its potential to further constrain the nPDFs. We find that $x_A$ derived from the reconstructed jets correlate well with the actual values of the $x$ probed in the nucleus and that at small-$x_A$ region the dijet production is dominated by the clean direct-photon contribution. Comparing the nPDF uncertainties with the expected statistical precision of the measurement at the LHC show that this observable could potentially provide precise nPDF constraints down to $x\sim 10^{-4}$.

\section*{Acknowledgments}
\vspace{-0.2cm}

\noindent The work has been supported by the Carl Zeiss Foundation and the Academy of Finland, Project 308301.

\end{document}